\documentclass[aps,prc,twocolumn,floatfix,superscriptaddress]{revtex4}

\usepackage{epsfig}
\usepackage{amsmath}

\usepackage{graphicx}

\begin{document}

\title{Enhancement of thermal photon production in event-by-event hydrodynamics}
\author{Rupa Chatterjee}
\affiliation{Department of Physics, P. O. Box 35 (YFL), FI-40014 University of Jyv\"askyl\"a, Finland}
\author{Hannu Holopainen}
\author{Thorsten Renk}
\author{Kari J. Eskola}
\affiliation{Department of Physics, P. O. Box 35 (YFL), FI-40014 University of Jyv\"askyl\"a, Finland}
\affiliation{Helsinki Institute of Physics, P.O.Box 64, FI-00014 University of Helsinki, Finland}

\begin{abstract}
Thermal photon emission is widely believed to reflect properties of the earliest, hottest evolution stage of the medium created in ultra-relativistic heavy-ion collisions. Previous computations of photon emission have been carried out using a hydrodynamical medium description with smooth, averaged initial conditions. Recently, more sophisticated hydrodynamical models which calculate observables by averaging over many evolutions with event-by-event fluctuating initial conditions (IC) have been developed. Given their direct connection to the early time dynamics, thermal photon emission appears an ideal observable to probe fluctuations in the medium initial state. In this work, we demonstrate that including fluctuations in the IC may lead to an enhancement of the thermal photon yield of about a factor of 2 in the region $2 < p_T < 4$ GeV/$c$ (where thermal photon production dominates the direct photon yield) compared to a scenario using smooth, averaged IC. Consequently, a much better agreement with PHENIX data is found. This can be understood in terms of the strong temperature dependence of thermal photon production, translating into a sensitivity to the presence of 'hotspots' in an event and thus establishing thermal photons as a suitable probe to characterize IC fluctuations.
\end{abstract}

%\pacs{25.75.-q,12.38.Mh}
\maketitle
\section{Introduction}
The recent observations of large elliptic flow~\cite{fl1, fl2} and energy loss of high energy partons in the medium~\cite{jet1, jet2} strongly suggest the formation of Quark-Gluon Plasma (QGP), a hot and dense state of quarks and gluons, in ultra-relativistic heavy ion collisions. The results from the Relativistic Heavy Ion Collider (RHIC) during the last decade as well as the recent results from the Large Hadron Collider (LHC)~\cite{alice} are consistent with theoretical calculations which assume early thermalization and a nearly ideal fluid behaviour of the system.

The evolution of the system from initial formation to final breakup is reflected in various different observables. Electromagnetic radiation has long been proposed as the  most promising and efficient probe to characterize the initial state of the hot and dense matter produced in the collisions~\cite{phot}. Unlike bulk hadrons, which are emitted from the freeze-out hyper-surface, photons come out from every phase of the expanding fireball, suffer negligible final state interactions (since electromagnetic coupling $\alpha \, \ll \,$ strong coupling $ \alpha_s $), and  carry undistorted information about the medium conditions at their production points. The thermal emission of photons from the evolving system shows a very strong temperature dependence: the high $p_T$ photons are emitted mostly from the hot and dense early stages of matter when the hydrodynamic flow is weak, whereas those at lower $p_T$ are emitted from the relatively cold later parts of the system where a large build-up of radial flow boosts their transverse momentum. Thermal photons with $p_T \ > \ 1 \ \rm{GeV/}$$c$, are therefore expected to provide a glimpse of the early part of the expansion history when the fireball is in the plasma phase, complementary to the spectra of bulk hadrons which reflect the late conditions close to decoupling. 

Initially direct photon spectra were studied mostly to extract the temperature of the system. In recent times, several other photon observables have shown significantly more potential to understand the dynamics of the system, for example, intensity interferometry of photons for space time information of the evolving system~\cite{interfero1,interfero2}, momentum anisotropy of the initial partons as well as formation time of QGP from elliptic flow of thermal photons~\cite{cfhs, cs}, study of jet-quenching  through photons produced from jet medium interaction~\cite{fms_phot, tgfh, photon_tagged} etc. In addition to that, a higher level of sophistication has been reached in the estimation of prompt photons from next to leading order (NLO) perturbative QCD calculations~\cite{pat1} where the result explains the direct photon spectrum from p+p collisions at RHIC quite well for $p_T > 1$ GeV/$c$~\cite{phenix1}. Also photons from jet-plasma interaction, investigated in detail recently~\cite{tgfh}, have been suggested as an important contribution
to the direct photon yield at $p_T > 2$ GeV/$c$.

Ideal hydrodynamics with a smooth initial density profile has been used successfully to model the evolution of the system at RHIC where the experimental data for hadron spectra and elliptic flow are well reproduced up to $p_T \sim 2$ GeV/$c$ (see e.g.~\cite{hydro,Niemi:2008ta}). However, for any given event there are inhomogeneities in the initial energy density profile and therefore event-by-event fluctuating density profiles are more realistic than a smooth initial  density distribution. Thus, the important question is whether these event-by-event fluctuating initial conditions (IC) have any sizable effect on the physical observables if they are computed by averaging over many final states rather than considering an averaged initial density profile.

It has been shown recently that hydrodynamics with event-by-event fluctuating IC reproduces the experimental charged particle elliptic flow even for the most central collisions at RHIC~\cite{hannu} which was underestimated by all earlier hydrodynamic calculations using smooth IC. Several other studies have shown that  fluctuating IC give a better agreement with the experimental charged particle spectra at high $p_T$ by increasing the number of particles there~\cite{hama, hannu}. These fluctuations may also help to understand the various structures observed in two-particle correlations~\cite{andrade, schenke}. Since thermal photons are sensitive to the initial temperature, they are especially suitable for probing fluctuations in the IC.  

 Thermal radiation is predicted to be the dominant source of direct photons in the range $1 \le p_T \le 3$ GeV/$c$~\cite{phenix1,dks_qm08}. Photon results using smooth profile and latest rates~\cite{trg} fall, however, well below the experimental data points (see Fig.~4 of Ref.~\cite{phenix1}) in this $p_T$ range. In the present work we report a substantial $p_T$ dependent  enhancement in the production of thermal photons for $p_T  > 1$ GeV/$c$ with an event-by-event fluctuating IC relative to a smooth initial-state averaged profile in the ideal hydrodynamic calculation. Consequently, our results with fluctuating density profile show a better agreement with the PHENIX experimental data for $2 < p_T < 4$ GeV/$c$. 
%%%% addition%%%%%%%%
However it is to be noted that the main aim of this paper is to investigate the effect of initial state fluctuations on the production of thermal photons rather than explaining the experimental direct photon spectrum, which contains significant contributions from other sources apart from thermal radiation. 
%%%%%%%%%%%%%%%%%%%
Also importantly, we find that the enhancement scales with the initial fluctuation size.

\section{Event-by-event hydrodynamics and initial density profile}
We use the event-by-event hydrodynamical model developed in~\cite{hannu}, which has been applied successfully to study the elliptic flow with fluctuating initial states. The standard two-parameter Woods-Saxon nuclear density profile is used to randomly distribute the nucleons in the colliding nuclei. Two nucleons $i$ and $j$ from different nuclei are then assumed to collide whenever they satisfy the relation
\begin{equation}
  (x_i - x_j)^2 + (y_i-y_j)^2 \le \frac{\sigma_{NN}}{\pi},
\end{equation}
where $\sigma_{NN} = 42 \ {\rm mb}$ is the inelastic nucleon-nucleon cross section at $\sqrt{s_{NN}} = 200 $ GeV, and $x_i,y_i$ denote the positions of the nucleons in the transverse plane.

The initial density profile is taken to be proportional to the number of wounded nucleons (WN), where  entropy density $s$ (or the energy density $\epsilon$) is distributed in the $(x,y)$ plane around the wounded nucleons using a 2D Gaussian as a size function,
\begin{equation}
  s[\epsilon] (x,y) = \frac{K}{2 \pi \sigma^2} \sum_{i=1}^{\ N_{\rm WN}} \exp \Big( -\frac{(x-x_i)^2+(y-y_i)^2}{2 \sigma^2} \Big).
 \label{eq:eps}
\end{equation}
We refer to this as sWN (eWN) profile. The parameter $K$ is a fixed overall normalization constant and $\sigma$ is a free  parameter determining the size of the fluctuations.

The effective interaction radius for the colliding nucleons, $\sqrt{\sigma_{NN}/\pi}/2 \ \sim$ 0.6~fm, sets a natural order of magnitude for the size parameter. Similarly to Ref.~\cite{hannu}, the default value of $\sigma$ is taken as 0.4 fm, however we test the sensitivity of the results to $\sigma$ by varying $\sigma$ from 0.4 to 1.0 fm. We consider the 0--20\% most central Au+Au collisions where the value of ${\rm N_{\rm WN}}$ fluctuates from 391 to 197 according to the Monte Carlo Glauber model. This corresponds to an average impact parameter of about 4.4 fm.
 
Using this procedure we thus find intra-event fluctuations (i.e. 'hotspots' and 'holes') in the IC in a given event, as well as inter-event fluctuations in the total entropy between any two events. We call events with larger than average entropy production "hot events" and those with smaller than average entropy production "cold events".

For Au+Au collisions at $\sqrt{s_{NN}} = 200 $ GeV the constant $K$ is taken as $102.1  \ \rm{fm^{-1}} \ (37.8 \ \rm {GeV}/\rm{fm})$ for the sWN (eWN) profile, which reproduces the initial total entropy of Ref.~\cite{Niemi:2008ta} when averaging over many initial states in central collisions for $\sigma=0.4 \rm{~fm}$. 
As in~\cite{hannu}, the initial time is fixed at $\tau_0 = 0.17$ fm/$c$ motivated by the EKRT minijet saturation model~\cite{Eskola:1999fc, Niemi:2008ta} for all the events. 

Assuming longitudinal boost invariance, we solve the (2+1) dimensional ideal hydrodynamic equations using the SHASTA algorithm \cite{Boris,Zalesak} which is able to propagate shock waves possible with the irregular IC. The Equation of State (EOS)
which shows a sharp crossover transition from plasma phase to hadronic matter is taken from~\cite{Laine:2006cp}. 
The temperature for freeze-out is taken as 160 MeV, which reproduces the measured $p_T$ spectrum of pions~\cite{Adler:2003cb} well for both the profiles eWN and sWN.

\section{Thermal photons}
The rate of thermal photon production in the QGP depends on the momentum distribution of the partons which are governed by the thermodynamical conditions of the matter.

The quark-gluon Compton scattering and quark-anti-quark annihilation are the leading order processes for photon production in the partonic phase, whereas in next to leading order, bremsstrahlung process contributes significantly to the production~\cite{amy}. In a hot hadronic gas at a temperature of the order of pion mass, $\pi$ and $\rho$ mesons contribute dominantly to the photon production because of the low mass of pions and the large spin iso-spin degeneracy of $\rho$ mesons, making them the most easily accessible particles in the medium~\cite{kls}. The leading photon producing channels involving  $\pi$ and $\rho$ mesons are $\pi \pi \ \rightarrow \ \rho \gamma$, $\pi \rho \ \rightarrow \ \pi \gamma$, and $\rho \ \rightarrow \ \pi \pi \gamma$.

We use the rates $R=EdN/d^3p d^4x$ of~\cite{amy} for the plasma and those of~\cite{trg} for the hadronic matter which at present can be considered as the state of the art. The transition from the plasma rates to the hadronic rates is assumed to happen at a temperature of 170 MeV in this study. 
The total thermal emission from the quark and hadronic matter phases are obtained by integrating the emission rates over the space-time history of the fireball as  
\begin{equation}
\label{eq1}
  E\, dN/d^3p = \int d^4x\, R\Big(E^*(x),T(x)\Big),
%  \bigl[(...)\exp(-p{\cdot}u(x)/T(x))\bigr]\,d^4x\,,
\end{equation}
where $E^*(x)=p^{\mu}u_{\mu}(x)$. The 4-momentum of the photon is 
 $p^\mu{\,=\,}(p_T \cosh Y, p_T\cos\phi,p_T\sin\phi,p_T \sinh Y)$, and the 4-velocity of the flow field is $u^\mu = \gamma_T\bigl(\cosh \eta,v_x,v_y,$ $\sinh\eta\bigr)$ with $\gamma_T = (1{-}v_T^2)^{-1/2}$, $v_T^2{\,=\,}v_x^2{+}v_y^2$. The volume element is $d^4x{\,=\,}\tau\, d\tau \, dx \, dy\, d\eta$, where $\tau{\,=\,}(t^2{-}z^2)^{1/2}$ is the longitudinal proper time and $\eta{\,=\,}\tanh^{-1}(z/t)$ is the space-time rapidity. The photon
momentum is parametrized by its rapidity $Y$, transverse momentum $p_T$, and azimuthal emission angle $\phi$. 
%--------------------------------------------------------------------------------------
%--------------------------- Fig1------------------------------------------------------
%--------------------------------------------------------------------------------------
\begin{figure}
\centerline{\includegraphics*[width=7.6cm]{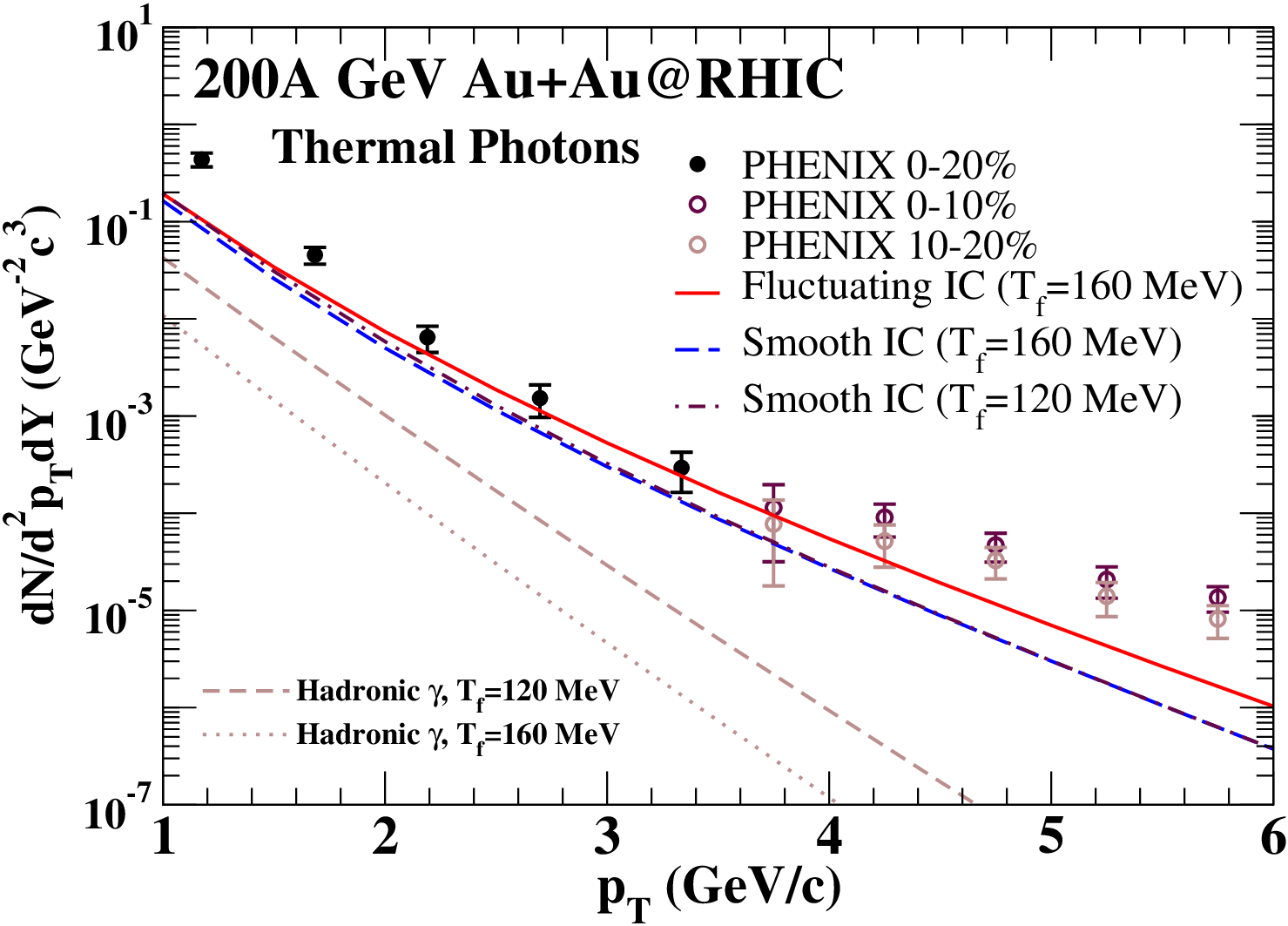}}
\caption{(Color online) Thermal photons from a smooth (long dashed blue line) and fluctuating (solid red line) sWN initial density profiles along with experimental direct photon data from PHENIX~\cite{phenix1,phenix2} for the 0--20\% centrality bin. The fluctuation size parameter $\sigma=0.4$~fm here. Also shown are the hadronic contributions for the smooth IC cases.}
\label{fig1}
\end{figure}
%--------------------------------------------------------------------------------------
%---------------------------End Fig1---------------------------------------------------
%--------------------------------------------------------------------------------------
\section{results}
We compare our results from smooth and fluctuating sWN initial density profiles with PHENIX data for 0--20\% centrality bin in  Fig.~\ref{fig1}. The smooth initial density profile here is obtained by taking an average of 1000 fluctuating initial profiles which is enough to remove essentially all fluctuations. To test whether there is a residual dependence on the IC averaging procedure we vary the value of $\sigma$ from 0.4 to 1.0 fm.  The thermal photon spectrum from the resulting smooth initial density profiles is almost independent of  $\sigma$ in the low $p_T$ region. Only at high $p_T$ ($> 4$ GeV/$c$), a change in the value of $\sigma$ from 0.4 to 1.0 decreases the photon production there by 20--30\%.

Photons from the fluctuating IC scenario are estimated by averaging photon spectra from sufficiently large number of random events. We make sure that addition of another very hot or very cold event does not change the results significantly.

We find that in the entire $p_T$ range the spectra are dominated by the QGP radiation both for the smooth as well as for the fluctuating IC. This is not due to our choice of large freeze-out temperature $T_f$ (160 MeV):
We have checked that a prolonged hydro evolution towards a smaller value like $T_f= $ 120 MeV contributes significantly below 1.5 GeV/$c$ but is not substantial for $p_T > $ 2 GeV/$c$, as shown by the brown dashed dotted line in Fig~\ref{fig1}.  Photons from (only) the hadronic phase are shown by dotted and short dashed gray lines for $T_f$ values of 160 and 120 MeV respectively for the smooth IC. Both these lines fall well below the plasma and the total contribution. It is to be noted that a bag model EOS produces more hadronic photons near the transition region compared to a lattice based EOS, however the contribution from the hadronic phase is not significant for $p_T > 2$ GeV/$c$ even for a bag model EOS. 
We also note that the Bag model yield of photons at high $p_T$ is about 50 \% smaller than the results shown in here. 
%--------------------------------------------------------------------------------------
%--------------------------- Fig2------------------------------------------------------
%--------------------------------------------------------------------------------------
\begin{figure}
\centerline{\includegraphics*[width=7.2cm]{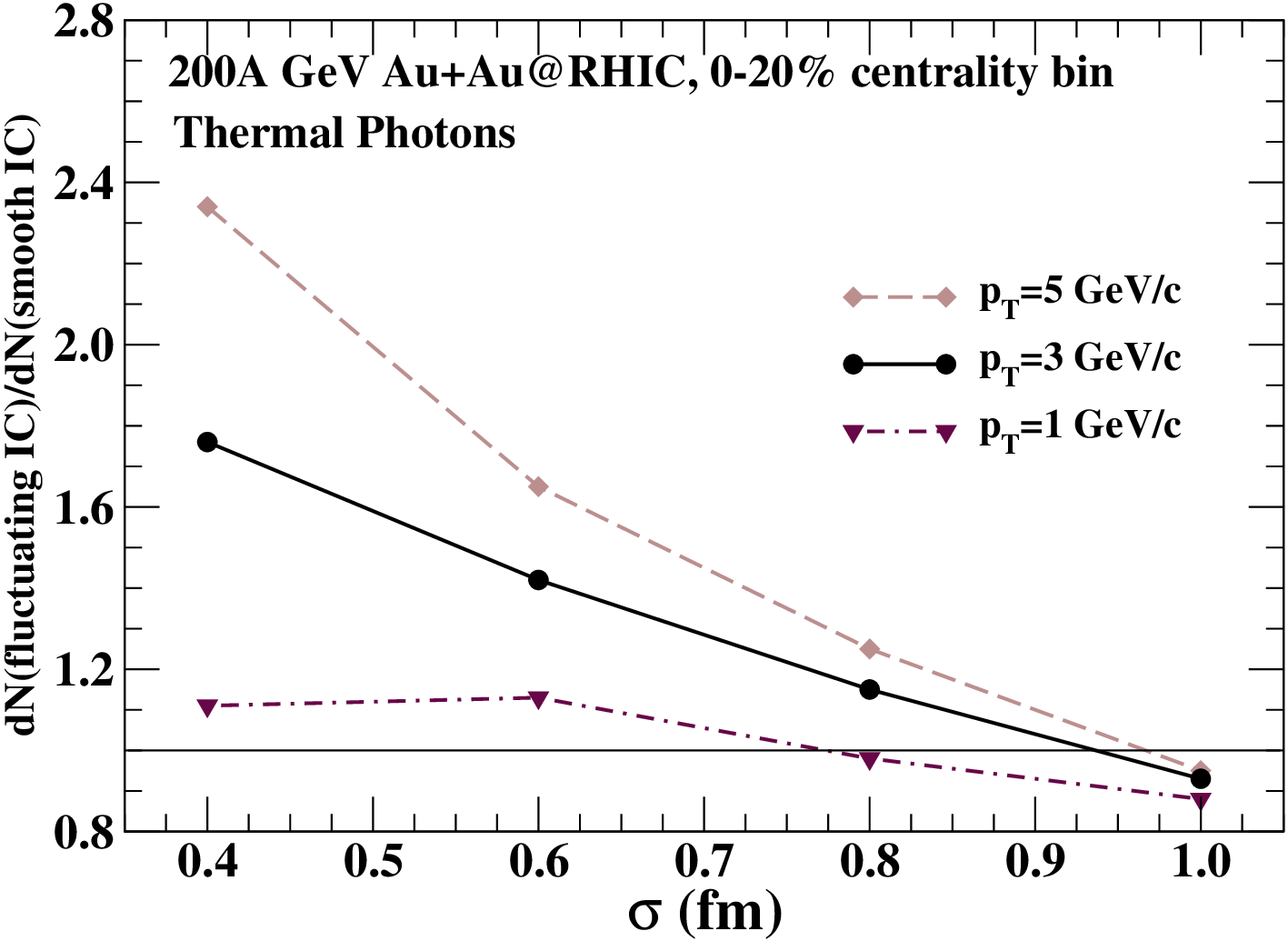}}
\caption{(Color online) Ratio of photon production from fluctuating and smooth IC at different $p_T$ as a function of the fluctuation size parameter $\sigma$.}
\label{fig2}
\end{figure}
%--------------------------------------------------------------------------------------
%--------------------------End Fig2----------------------------------------------------
%--------------------------------------------------------------------------------------

%%%%%%%%%%%%%%%%%%%%%%%%%%%%%%%%%%%%%%% modified %%%%%%%%%%%%%%%%%%%
The slope of the photon spectrum from the fluctuating IC is about 10\% flatter compared to the slope from the smooth one in the region $2 \le p_T \le 4$ GeV/$c$ and  the two results differ almost by a factor of 2 (see Fig.~\ref{fig2}) for $p_T > 2$ GeV/$c$.
%%%%%%%%%%%%%%%%%%%%%%%%%%%%%%%%%%%%%%%%%%%%%%%%%
 In the region $p_T < 2$ GeV/$c$, the fluctuating case produces 20--40\% more photons than the smooth profile. The thermal emission of photons is exponential in temperature and linear in radiating volume. 
As a result, the hotspots in the fluctuating events produce more high $p_T$ photons compared to the smooth profile. On the other hand, the low $p_T$ part of the spectrum comes from the relatively cold more volume-dominated  later plasma stage, where the emission is not affected significantly due to the change in the IC. Thus the difference between the two profiles is small near $p_T\sim 1$ GeV/$c$ and enhanced towards higher $p_T$. 

%%%%%%%%%%%% added %%%%%%%%%%%%%%%%%%
It may be noted that a similar enhancement of the thermal photon production by a factor $\sim 2$ in the region $1<p_T<4$ GeV/$c$ using the same smooth IC can also be obtained by choosing a smaller initial time $\tau_0$. In this case, one needs to reduce the the value of $\tau_0$ to half of its original value~\cite{cs}.  A systematic study of thermal photons at different centrality bins has the potential to distinguish between the effects of fluctuations in the initial state and the formation time.  

Similar to the photon results, hardening of the spectra  is observed also for hadrons where the fluctuation-driven radial flow makes their spectra harder compared to  the spectra obtained from a smooth IC. This is because the hadrons are emitted from the much later stage of the system expansion, i.e. the surface of freeze-out~\cite{hannu,uli1}. For photons, the difference between the fluctuating and the smooth IC is an early time effect (due to the presence of hotspots) when the radial flow is still very small. Although the photons from the late hadronic phase are affected significantly due to the increased radial flow in the fluctuating IC, the sum spectrum, which is dominated by plasma radiation in the entire $p_T$ range, remains almost unaffected.
%%%%%%%%%%%%%%%%%%%%%%%%%%%%%%%%%%%%%%%%%%%%%

Thus, the production of thermal photons shows a very strong dependence on the averaging procedure.  We observe that the results for fluctuating IC show better agreement with the experimental data compared to earlier studies with smooth initial profile while leaving enough space for the prompt contribution and jet conversion photons. This may also affect the photon elliptic flow results~\cite{cfhs}.

The sensitivity of the results to the fluctuation size parameter $\sigma$ is shown in Fig.~\ref{fig2}, where the ratio of the results from fluctuating and smooth IC are plotted as a function of $\sigma$ for different values of $p_T$. The enhancement is maximal for higher values of $p_T$ and lower values of $\sigma$ which clearly shows the dominance of the hotspots in high $p_T$ photon production. For $\sigma \ge 0.8$ fm,  profiles with fluctuating IC do not show pronounced differences to the smooth case as the ratio approaches 1.

%--------------------------------------------------------------------------------------
%--------------------------- Fig3------------------------------------------------------
%--------------------------------------------------------------------------------------
\begin{figure}
\centerline{\includegraphics*[width=7.2cm]{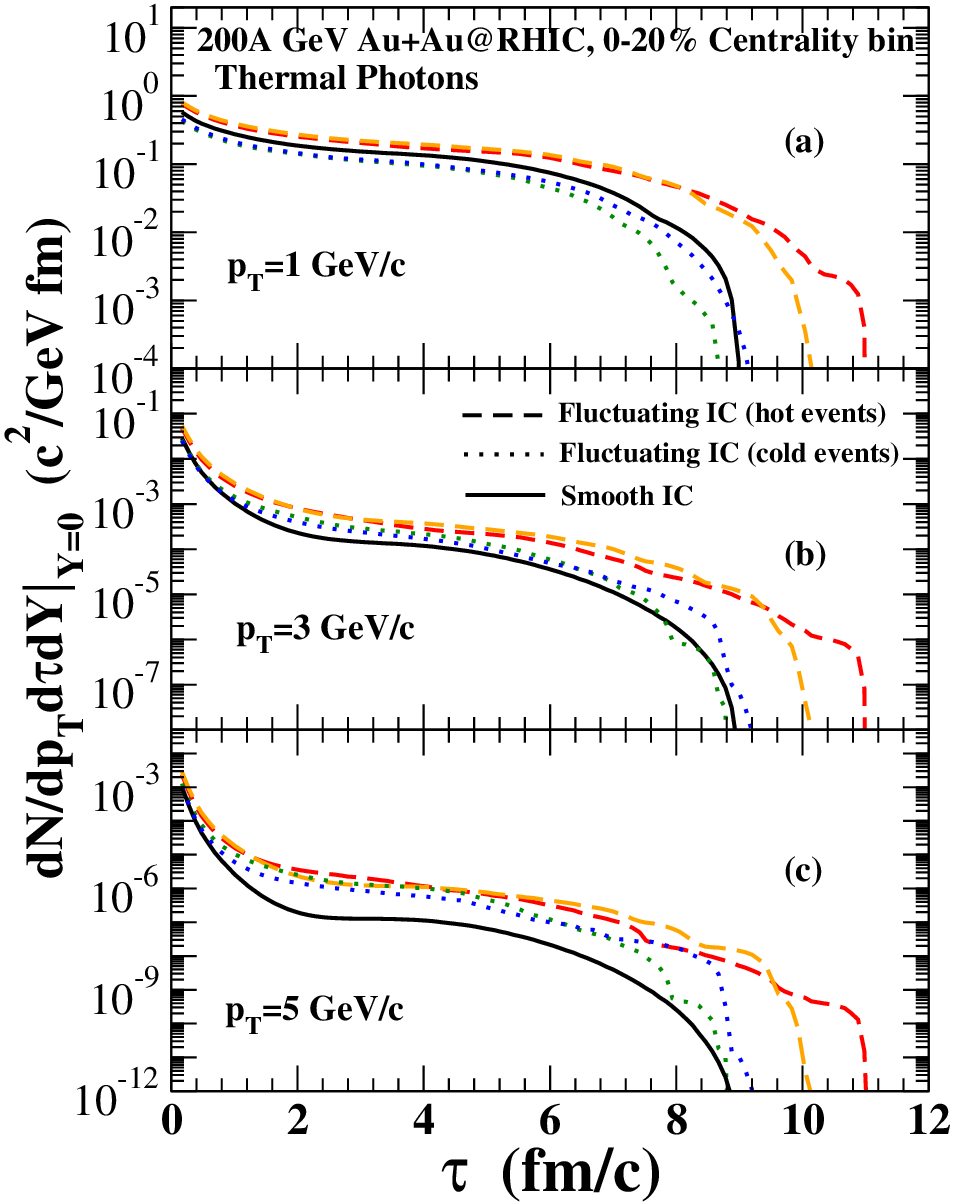}}
\caption{(Color online)Time evolution of thermal photons for transverse momentum values of (a) 1 GeV/$c$, (b) 3 GeV/$c$, and (c) 5 GeV/$c$. Results are compared with  an initial state averaged event and four different random events.}
\label{fig3}
\end{figure}
%--------------------------------------------------------------------------------------
%--------------------------- End Fig3--------------------------------------------------
%-------------------------------------------------------------------------------------- 

We also notice that the photon results are quite sensitive to the initial profiles; the results for sWN are harder than those with the eWN. The entropy initialization leads to more pronounced maxima in the energy density distribution (causing hotter initial state and harder spectra) than their energy initialization counterparts which was observed in some earlier studies also~\cite{kolb}. However, the photon spectrum  from fluctuating IC considering the eWN profile shows a similar enhancement relative to the smooth profile as observed for sWN in Fig.~\ref{fig1}.

In Fig.~\ref{fig3} the time evolution of the photon results for smooth (solid lines) and fluctuating (dashed lines) IC are shown at $p_T$ values of 1, 3, and 5 GeV/$c$, to illustrate the dynamics leading  to the results presented in Fig.~\ref{fig1}. For the fluctuating IC we choose two very hot events (red and orange dashed lines) and two relatively cold events (blue and green dotted lines) whereas the smooth IC event is shown by the solid black lines in the figure. The $dN/dp_Td\tau dY$ results clearly demonstrate that most of the high $p_T$ photons are emitted quite early where the photon emission rate drops by two orders of magnitude within a time period of about 2 fm/$c$ for $p_T \ge 3$ GeV/$c$. The cold events reflect a relatively smaller system lifetime. The most interesting observation is that the cold events produce more photons at $p_T \ge 3$ GeV/$c$ compared to the smooth profile due to the presence of a few hotspots.

\section{Summary and conclusions}
In conclusion, we find a large enhancement in the production of  thermal photons for event-by-event fluctuating IC compared to a smooth profile in the ideal hydrodynamic calculation. For $p_T < 2 $ GeV/$c$, the production is increased by 20--40\% for the fluctuating initial conditions, whereas for $p_T > 2$ GeV/$c$, it increases by about a factor of 2, which gives better agreement with PHENIX data in the region $2 < p_T < 4$ GeV/$c$. 
%%%%%%%%%%%%addition
 A small value of the plasma formation time (0.17 fm/$c$) ensures that a fraction of the pre-equilibrium photons is accounted for which is an important source of direct photons at large values of transverse momentum and hard to estimate separately. The contributions from the initial hard processes and jet-matter interactions are significant for $p_T > 2$ GeV/$c$, and should be added with the thermal radiation in order to explain the data in the entire $p_T$ range shown in Fig.~\ref{fig1}.
%%%%%%%%%%%%%%%%%%%
This enhancement is a result of the strong temperature dependence of the thermal photon emission rates which specifically probe the hotspots. 
 We checked that the photon spectra for the sWN profile are harder than the spectra for eWN profile, however the quantitative difference between the fluctuating and smooth IC remains similar for both the cases.  A more detailed investigation of the wounded nucleons and binary collision profiles for both entropy and energy initializations combined with photon elliptic flow from the fluctuating IC would be very valuable. The difference between the smooth and fluctuating IC is expected to increase for peripheral collisions as well as for lower collision energies. We postpone these issues for a future study. 

\begin{acknowledgments} 
We gratefully acknowledge financial support by the Academy of Finland. TR and RC are supported by the Academy researcher program (project  130472) and KJE by a research grant (project 133005). In addition, HH is supported by the national Graduate School of Particle and Nuclear Physics. We would like to thank U. Heinz for useful discussions. 
\end{acknowledgments}


\begin{thebibliography}{99}
\bibitem{fl1} C.~Adler {\it et al.} [STAR Collaboration], Phys.\ Rev.
\ Lett. {\bf 87}, 182301 (2001); ibid {\bf 89}, 132301 (2002); ibid
{\bf 90}, 032301 (2003); S.~S.~Adler {\it et al.} [PHENIX Collaboration],
Phys.\ Rev. \ Lett. {\bf 91}, 182301 (2003).


\bibitem{fl2} P.~Huovinen, P.~Kolb, U.~Heinz, P.~V.~Ruuskanen, and
S.~Voloshin, \ Phys.\ Lett. \ B {\bf 503}, 58 (2001); D.~Teaney, J.~Lauret,
 and E.~Shuryak, nucl-th/0110037.

\bibitem{jet1} X.~N.~Wang, Phys.\ Rev. \ C {\bf 63}, 054902 (2001);
M.~Gyulassy, I.~Vitev, and X.~N.~Wang, Phys.\ Rev. \ Lett. {\bf 86},
 2537 (2001).

\bibitem{jet2} K.~Adcox {\it et al.} [PHENIX Collaboration], Phys.
\ Rev.\ Lett. {\bf 88}, 022301 (2002); J.~Adams {\it et al.}
[STAR Collaboration], Phys.\  Rev.\ Lett. {\bf 91}, 172302 (2003).

\bibitem{alice} K. Aamodt {\it et al.} [ALICE Collaboration], Phys. \ Rev. \ Lett. {\bf 105}, 252301 (2010); arXiv:1011.3914.

\bibitem{phot} 
  P.~V.~Ruuskanen,
%``Electromagnetic probes of quark - gluon plasma in relativistic heavy ion
  %collisions,''
  Nucl.\ Phys.\  A {\bf 544}, 169 (1992), and references therein.


\bibitem{interfero1} S.~A.~Bass, B.~M\"uller, and D.~K.~Srivastava,
Phys. \ Rev. \ Lett. {\bf 93}, 162301 (2004).

\bibitem{interfero2} D. K. Srivastava and R. Chatterjee, Phys. \ Rev. \ C {80}, 054914 (2009); {\it Erratum-ibid.} C{\bf 81}, 029901 (2010); S. De, D. K. Srivastava, and R. Chatterjee, J. \ Phys. \ G {\bf 37}, 115004 (2010).


\bibitem{cfhs}R.~Chatterjee, E.~S.~Frodermann, U.~W.~Heinz and D.~K.~Srivastava,
  Phys.\ Rev.\ Lett.\  {\bf 96}, 202302 (2006).


\bibitem{cs} R. Chatterjee and D. K. Srivastava, Phys. \ Rev. \ C {\bf 79}, 021901(R) (2009); R. Chatterjee and D. K. Srivastava, Nucl. \ Phys. {\bf A830}, 503c (2009). 


\bibitem{photon_tagged} 
X.~N.~Wang, Z.~Huang and I.~Sarcevic,
  %``Jet quenching in the opposite direction of a tagged photon in  high-energy
  %heavy-ion collisions,''
  Phys.\ Rev.\ Lett.\  {\bf 77}, 231 (1996)
  [arXiv:hep-ph/9605213];
  %%CITATION = PRLTA,77,231;%%
 T. Renk, Phys. \ Rev. \ C {74}, 034906 (2006).

\bibitem{fms_phot} R.~J.~Fries, B.~M\"uller and D.~K.~Srivastava,
\ Phys. \ Rev. \ Lett. {\bf 90}, 132301 (2003); R.~J.~Fries, B.~M\"uller,
and D.~K.~Srivastava, \ Phys. \ Rev. \ C {\bf 72} 041902 (R) (2005).


\bibitem{tgfh} S. Turbide, C. Gale, S. Jeon, and G. D. Moore, Phys. \ Rev. \ C {\bf 72}, 014906 (2005); S.~Turbide, C.~Gale, E.~Frodermann, U.~Heinz
\ Phys.\ Rev.\ C {\bf 77}, 024909 (2008).


\bibitem{pat1} P. Aurenche, M. Fontannaz, J.-P. Guillet, B. A. Kniehl,
E. Pilon, and M. Werlen, \ Eur. \ Phys. \ J. \ C {\bf 9}, 107 (1999);  P.~Aurenche, M.~Fontannaz, J.~P.~Guillet, E.~Pilon, and M.~Werlen,
  Phys.\ Rev.\  D {\bf 73}, 094007 (2006).

\bibitem{phenix1} A. Adare {\it et al.} [PHENIX Collaboration], Phys.  \ Rev. \ Lett. {\bf 104}, 132301 (2010).

\bibitem{hydro}  K.~J.~Eskola, H.~Honkanen, H.~Niemi, P.~V.~Ruuskanen and S.~S.~Rasanen,
%``RHIC-tested predictions for low-p($T$) and high-p($T$) hadron spectra in
%nearly central Pb + Pb collisions at the CERN LHC,''
Phys.\ Rev.\  C {\bf 72}, 044904 (2005);
% H.~Niemi, K.~J.~Eskola and P.~V.~Ruuskanen,
%``Elliptic flow in nuclear collisions at the Large Hadron Collider,''
%Phys.\ Rev.\  C {\bf 79}, 024903 (2009);
  P.~Huovinen and P.~V.~Ruuskanen,
%``Hydrodynamic Models for Heavy Ion Collisions,''
 Ann.\ Rev.\ Nucl.\ Part.\ Sci.\  {\bf 56}, 163 (2006);
   C.~Nonaka and S.~A.~Bass,
%``Space-time evolution of bulk QCD matter,''
 Phys.\ Rev.\  C {\bf 75}, 014902 (2007).

\bibitem{Niemi:2008ta}
H.~Niemi, K.~J. Eskola, and P.~V. Ruuskanen,
\newblock Phys. Rev. {\bf C79}, 024903 (2009), 0806.1116.
  

\bibitem{hannu} H. Holopainen, H. Niemi, and K. Eskola, Phys. \ Rev. \ C {\bf 83}, 034901 (2011). 

\bibitem{hama}
Y.~Hama, T.~Kodama, and O.~Socolowski, Jr.,
\newblock Braz. J. Phys. {\bf 35}, 24 (2005), hep-ph/0407264.

\bibitem{andrade} R.~Andrade, F.~Grassi, Y.~Hama, T.~Kodama, and O.~Socolowski, Jr., Phys. Rev. Lett. {\bf 97}, 202302 (2006), nucl-th/0608067;
R.~P.~G. Andrade, F.~Grassi, Y.~Hama, T.~Kodama, and W.~L. Qian, Phys. Rev. Lett. {\bf 101}, 112301 (2008), 0805.0018.


\bibitem{schenke} B.~Schenke, C.~Gale and S.~Jeon,
 Phys.\ Rev.\  C {\bf 80}, 054913 (2009).

\bibitem{dks_qm08} D. K. Srivastava, \ J. \ Phys. \ G {\bf 35}, 104026 (2008),
 arXiv:0805.3401.

\bibitem{trg} S.~Turbide, R.~Rapp, and C.~Gale, \ Phys. \ Rev. \ C {\bf 69},
 014903 (2004).

\bibitem{Eskola:1999fc}
K.~J. Eskola, K.~Kajantie, P.~V. Ruuskanen, and K.~Tuominen,
\newblock Nucl. Phys. {\bf B570}, 379 (2000), hep-ph/9909456.

\bibitem{Boris}
J.~P. Boris and D.~L. Book,
\newblock J. Comput. Phys. {\bf A11}, 38 (1973).

\bibitem{Zalesak}
S.~T. Zalesak,
\newblock J. Comput. Phys. {\bf A31}, 335 (1979).

\bibitem{Laine:2006cp}
M.~Laine and Y.~Schroder,
\newblock Phys. Rev. {\bf D73}, 085009 (2006), hep-ph/0603048.

\bibitem{Adler:2003cb}
S.~S. Adler {\em et~al.},
\newblock Phys. Rev. {\bf C69}, 034909 (2004), nucl-ex/0307022.


\bibitem{amy} P.~Arnold, G.~D.~Moore, and L.~G.~Yaffe, JHEP {\bf 0112}, 009
(2001).

\bibitem{kls} J.~I.~Kapusta, P.~Lichard and D.~Seibert,
  Phys.\ Rev.\  D {\bf 44}, 2774 (1991)
  [Erratum-ibid.\  D {\bf 47}, 4171 (1993)]
  [Phys.\ Rev.\  D {\bf 47}, 4171 (1993)].
 

\bibitem{phenix2} S. S. Adler {\it et al.} [PHENIX Collaboration], Phys. \ Rev. \ Lett. {\bf 94}, 232301 (2005).

\bibitem{uli1} Z. Qiu and U. W. Heinz, arXiv:1104.0650. 
    

\bibitem{kolb}
  P.~F.~Kolb and U.~W.~Heinz, Hydrodynamic description of ultrarelativistic heavy-ion collisions, in {\it Quark-Gluon Plasma 3}, edited by R. C. Hwa and X. N. Wang, p. 634, World Scientific, Singapore, 2004,
  arXiv:nucl-th/0305084.

\end{thebibliography}
\end{document}